
\input phyzzx
\overfullrule=0pt

\def\dAlemb{\hbox{$\sqcap\kern-0.67em\sqcup$}}
\def\dslash{\not{\hbox{\kern-2pt $\partial$}}}
\def\Qslash{\vert{\hbox{\kern-5pt Q}}}
\def\Rslash{\vert{\hbox{\kern-5.5pt R}}}
\def\hslash{\not{\hbox{\kern+1.5pt h}}}
\def \f{f^a_{bc}}
\def \fcr{\f \c b {} \r c {} }

\def\({\left(}           \def\){\right)}
\def\[{\left\lbrack}           \def\]{\right\rbrack}
\def\{{\left\lbrace}           \def\}{\right\rbrace}
\def\co{cohomology\ }
\def\G{${G\over G}\ $}

\def\c#1#2{\chi_{#1}^{#2}}

\def\r#1#2{\rho_{#1}^{#2}}
\def\jt#1#2{{J_{(tot)}}_{#1}^{#2}}

\def\t{\tilde}
\def\pa{\partial}
\def \f{f^a_{bc}}
\def \fcr{\f \c b {} \r c {} }

\def \pa{\partial}

\def\W{$W_N\ $ }
\def\G{${G\over G}\ $}
\def\GH{${G\over H}\ $}

 \def\t{\tilde}
 \def\pa{\partial}
\def\bpa{\bar\partial}

\def\Q{$QCD_2$}

\def\Cf{\varphi}
\def\snc{$SU(N_C)$}
\def\cmp#1{{\it Comm. Math. Phys.} {\bf #1}}

\def\pl#1{{\it Phys. Lett.} {\bf #1B}}

\def\prd#1{{\it Phys. Rev.} {\bf D#1}}

\def\np#1{{\it Nucl. Phys.} {\bf B#1}}

\def\jmath#1{{\it J. Math. Phys.} {\bf #1}}
\def\mpl#1{{\it Mod. Phys. Lett.} {\bf A#1}}

\REF\Gross{David J. Gross, ``Two Dimensional QCD as a string theory,''
hep-th/9212149, \np {400} 161 (1993).}
\REF\DFS{G. D. Date, Y. Frishman, and J. Sonnenschein,
{\it Nucl. Phys.}\  {\bf B283}  (1987) 365.}
\REF\FS {Y. Frishman and J. Sonnenschein, \np  {294} (1987) 801 ;
``Bosonization and QCD in Two Dimensions,''
hep-th/9207017, {\it Physics Reports} {\bf 223 \# 6}  (1993)
309.}
\REF\BaRS{K. Bardacki, E. Rabinovici and B. Saering, \np {299} (1988) 151.}
\REF\GK{K. Gawedzki and A. Kupianen, \pl {215} (1988) 119;
\np {320} (1989) 649.}
\REF\KS{D. Karabali and H. J. Schnitzer, \np {329} (1990) 649.}
\REF\Migdal{A. A. Migdal {\it  Sov. Phys. JETP. }{\bf 42} (1975) 413.}
\REF\Rus{B. Rusakov, \mpl {5} (1990) 693.}
\REF\WQC{E. Witten, \cmp {141} (1991) 153.}
\REF\Kut{D. Kutasov, \pl {233} (1989) 369.}
\REF\Pat{A. Patrascioiu, \prd {15} (1977) 3592.}
\REF\PW{A. Polyakov and P. B. Wiegmann, \pl {131} (1983)  121.}
\REF\us{O. Aharony, O. Ganor, N. Sochen, J. Sonnenschein and S.
Yankielowicz, ``Physical States in G/G Models and 2d Gravity,''
hep-th/9204095, \np {399} (1993) 527.}
\REF\MiPo{J. A. Minahan and A. P. Polychronakos,
``Equivalence of Two Dimensional QCD and the $c=1$ Matrix Model,''
hep-th/9303153, \pl {312} (1993) 155.}
\REF\uss  { O. Aharony, J. Sonnenschein and S.Yankielowicz,
``G/G Models and \W  strings,''
hep-th/9206063, \pl {289}  (1992) 309.}
\REF\usss  {O. Aharony, O. Ganor, J. Sonnenschein and S.~Yankielowicz,
``On the twisted G/H topological models,''
hep-th/9208040, \np {399} (1993) 560.}
\REF\FHK{J.~Ellis, Y.~Frishman, A.~Hanany and M.~Karliner,
``Quark solitons as constituents of hadrons,''
hep-ph/9204212, Nucl.\ Phys.\ B{\bf 382} (1992) 189.}
\REF\IsRa{J. Isidro and A. V. Ramallo,
``gl(N,N) current algebras and TFTs,''
hep-th/9307037, US-FT-3/93.}
\REF\Ber  { J. Sonnenschein and S. Yankielowicz, ``Non-Critical String
Models as Topological Coset Models,''
hep-th/9310077, TAUP-2098-94.}

\titlepage
\baselineskip=20pt
\rightline {WIS-93/108/Oct-PH}
\rightline{TAUP- 2093-93}
\title{ Subtleties in  QCD theory in Two
Dimensions} \author {Y. Frishman \footnote{*}
{Supported in part by the Israel Academy of Sciences.}
and A. Hanany}
\address{Department of Particle Physics \break
Weizmann Institute of Science \break
 76100 Rehovot Israel}
\author { J.~Sonnenschein
\footnote{\dagger}{Work supported in part
by the US-Israel Binational Science Foundation and the Israel
Academy of Sciences.}}
\address{ School of Physics and
Astronomy\break Beverly and Raymond Sackler \break
Faculty    of Exact Sciences\break
Ramat Aviv Tel-Aviv, 69987, Israel}
\abstract{It is shown that in a formulation of Yang-Mills theory in two
dimensions in terms of $A=if^{-1}\pa f$, $\bar A=i\bar f\bpa\bar f^{-1}$
with $f(z,\bar z)$, $\bar f(z,\bar z)\in[SU(N_C)]^c$  the complexification
of \snc  , reveals certain subtleties.
``Physical" massive color singlet states seem to exist.
When coupled to $N_F$ quarks the coupling constant is renormalized
in such a way that it vanishes for the pure Yang- Mills case. This
renders the above
states massless and unphysical.
In the abelian case, on the other hand,  the known results of the
Schwinger model are reproduced with no need of such a renormalization.
 The   massless \Q\
theory is analyzed in similar terms and
peculiar  massive states  appear, with a mass  of
$e_c\sqrt {N_F \over 2\pi}$.}

\chapter{Introduction}
 Pure Yang-Mills theory on compact Riemann surfaces
attracted recently much attention, mainly in relation to a possible
underlying string theory.\refmark{\Gross}

The structure of the  bosonized non-abelian massless \Q\refmark{\DFS,\FS}
is that of a gauged WZW model with an additional $F^2$ term of the gauge
fields. Apart from the pure gauge term, this is thus a special form of a
two dimensional coset model\refmark{\BaRS,\GK,\KS}. It was found that
translating the gauge fields into scalars $f$ and $\bar f$ via $A=if^{-1}
\pa f$, $\bar A=i\bar f\bpa\bar f^{-1}$ with $f(z,\bar z),\bar f(z,\bar
z)\in H^c$, the complexification of $H\equiv SU(N_C)$ leads to a convenient
formulation of the model.  The main advantage of this approach is that one
can then easily decouple the ``matter" and the gauge degrees of freedom.

In the present paper we point out that the $F^2$ term   requires a special
 treatment.
 The formulation of  pure
YM theory in terms of the $f$ variables seems naively to
 contain unexpected ``physical" massive color singlet states.
This result is obviously neither in accordance with our ideas of the
degrees of freedom of the model nor with the lattice\refmark{\Migdal,\Rus} and
continuum  solution of the theory.\refmark\WQC\
We show that similar ``naive" manipulations in the case  of $QED_2$ do
 reproduce the Schwinger model results.
Using  a coupling constant renormalization  introduced  by D.
Kutasov,\refmark\Kut
we show that in the limit of no matter degrees of freedom the
coupling constant is renormalized to  zero. In this case the unexpected
states
turn into massless ``BRST" exact states.
In the flavored \Q\
case a similar analysis shows the existence of
``physical" flavorless states of mass  $m^2={N_F\over 2\pi} e_c^2$.
It seems that
these states that could not be seen in  the analysis of `t Hooft,
 correspond to  the massive states of ref.[\Pat]

The paper is organized as follows. In section 2 we introduce the
formulation of \Q\ in terms of $A=if^{-1}\pa f$, $\bar A=i\bar f\bpa\bar
f^{-1}$ with $f(z,\bar z),\bar f(z,\bar z)\in[SU(N_C)]^c$, the
complexification of \snc. The gauge invariant and gauge fixed actions
are written down. The special case of pure YM theory is introduced in
section 3, where the equations of motion, the symmetries and the
corresponding currents are discussed. In section 4 we digress to analyze
the Schwinger model as the abelian analog of the formulation that was
introduced earlier for \Q. It is shown that a BRST quantization of the
model reproduces the well known spectrum of the model. The pure
non-abelian gauge theory on the plane is further discussed in section 5.
It is analyzed as a ``perturbed" topological \G\ model. We show that
without the introduction of renormalization the ``naive"
space of physical states include unexpected massive color singlets.
An alternative formulation of the Yang Mills theory
in terms of an auxiliary field  is analyzed
in section 6.
In section 7 the resolution of the mysterious states is presented.
Starting from the \Q\
case we show that in the $N_F\rightarrow 0$ limit the
unexpected states become massless and
indeed decouple.
Section 8 is devoted to a brief discussion  of the massless \Q\ model.
Again this is done using a  \G  interpretation.
A summary  and a brief  comparison with other methods of treating
YM and \Q\ is presented in section 9.

\chapter{ The action }
The bosonized version of \Q\ was shown
\refmark\DFS\ to be described
by the following action
$$\eqalign{S_{QCD_2}=&S_1(u)-{1\over
2\pi} \int d^2 z Tr(iu^{-1}\pa u \bar A + iu\bar \pa
u^{-1} A + \bar A u^{-1}A u-A \bar A )\cr
+&{m^2\over{2\pi}}\int d^2z :Tr_G[u + u^{-1}]:
+{1\over e_c^2}\int d^2 zTr_H\[F^2\]\cr}\eqn\mishwzw$$
where $u\in U(N_F\times N_C)$; $S_k(u)$ is a level $k$ WZW model;
$$S_k(u)={k\over 8\pi}\int d^2xTr(\partial_\mu u\partial^\mu u^{-1})
+ {k\over12\pi}\int_B d^3y\varepsilon^{ijk}Tr(u^{-1}\partial_iu)
(u^{-1}\partial_ju)(u^{-1}\partial_ku) \eqn{\mishcb}$$
$A$ and
$\bar A$ take their values in the algebra of $H\equiv SU(N_C)$; $F=\bar\pa
A-\pa\bar A+i[A,\bar A]$; $m^2$ equals $m_q\mu C$, where $\mu$ is the
normal ordering mass and $C={1\over2}e^\gamma$ with $\gamma$ Euler's
constant. Apart from the  last two terms which correspond to the quark
mass term and the YM term the rest of the action is a  level one
\GH coset model
with $G=U(N_F\times N_C)$.

We now introduce the following parameterization for the gauge fields
$A=if^{-1}\pa f,\bar A=i\bar f\bpa\bar f^{-1}$ with $f(z,\bar z),\bar
f(z,\bar z)\in SU(N_C)^c$. These type of variables were used frequently
in dealing with gauged WZW actions, for instance in computing the
effective action of \Q\refmark\PW\ and in the \G\ models.\refmark\us\
They may be interpreted as Wilson lines along the $z$ and $\bar z$
directions. \refmark{\MiPo} The gauged WZW part of the action, first
line of \mishwzw, then\refmark{\GK,\KS,\PW} takes the form
$$S_1(u,A) =S_1(fu\bar f) -S_1(f\bar f) \eqn\mishwzwh$$
The Jacobian of the change of variables from $A$ to $f$ introduces
\refmark{\GK,\KS} a dimension $(1,0)$ system of anticommuting ghosts
$(\rho,\chi)$ in the adjoint representation of $H$. The WZW part of the
action thus becomes
$$S_1(u,A) =S_1(fu\bar f) -S_1(f\bar f) +{i\over {2\pi}}\int d^2z
Tr_H[\rho\bar D \chi + \bar\rho D \bar\chi] \eqn\mishwzwh$$ where
$D\chi=\pa\chi-i[A,\chi]$. Our integration variables in the functional
integral are $if^{-1}df$ and $i\bar fd\bar f^{-1}$. This action involves
an interaction term of the
form $Tr_H(\bar\rho [f^{-1}\pa f,\bar\chi])$ and a similar term for
$\rho,\chi$. By performing a chiral rotation $\bar\rho\rightarrow
f^{-1}\bar\rho f$ and $\bar\chi\rightarrow f^{-1}\bar\chi f$ with
$\rho\rightarrow\bar f\rho\bar f^{-1}$ and $\chi\rightarrow\bar
f\chi\bar f^{-1}$, one achieves a decoupling
of the whole ghost system. The price of that is an additional
$S^{(H)}_{-2N_C}(f\bar f)$ term in the action (here trace over H only)
resulting from the
corresponding anomaly. This result can be derived by using a non-abelian
bosonization of the ghost system.\refmark{\usss} In this language the
ghost action takes the form
$$S_{gh}= S_{N_c}(l_1, A,\bar A) + S_{N_c}(l_2, A,\bar A) +
S_{twist}(l_1) + S_{twist}(l_2)
\eqn\mishsgh$$
where $l_1$ and $l_2$ are in the adjoint representation and $S_{twist}$
is a twist term given in ref. [\usss]. Now using the Polyakov-Wiegmann
formula \refmark{\PW} we get $$\eqalign{S_{gh}^b&=
S_{N_c}(fl_1\bar f)
- S_{N_c}(f\bar f) +S_{twist}(l_1)
+ S_{N_c}(fl_2\bar f) - S_{N_c}(f\bar f)  +S_{twist}(l_2) \cr&=
- S_{2N_c}(f\bar f)
+{i\over{2\pi}}\int d^2z\Tr_H[\bar\rho'\pa\bar\chi'
+\rho'\bar\pa\chi'] \cr}
\eqn\mishsghm$$
where  the last line has been transferred back to the ghost language.
Notice that unlike the ghost fields in \mishwzwh\ the new ghost fields
$\rho'$ and $\chi'$ are now gauge invariant.
It is interesting to note that the action given in \mishwzwh\
is non-local in terms of the local degrees of freedom $A$ and $\bar A$.
Note that had we done the right and left rotations separately,
we would have gotten $S_{-2N_c}(f)+S_{-2N_c}(\bar f)$, which however is
not vector gauge invariant, but rather a left-right symmetric scheme.

The full gauge invariant action including the anomaly contribution of the
anti-commuting part now reads
$$\eqalign{S_{QCD_2}=&S_1(u)+{1\over e_c^2}\int d^2z\Tr_H\[F^2\]
+{m^2\over{2\pi}}\int d^2z:\Tr_G\[f^{-1}u\bar f^{-1}+\bar fu^{-1}f\]:\cr
+&
\[S^{(H)}_{-(N_F+2N_C) }(f\bar f)
+{i\over{2\pi}}\int d^2z\Tr_H\[\bar\rho'\pa\bar\chi'+\rho'\bar\pa\chi'\]\]
\cr}\eqn\mishwzwh$$
In deriving equation \mishwzwh\ we used a redefinition $fu\bar
f\rightarrow u$. This does not require an extra determinant factor
\refmark{\GK,\KS}. Also, as $S^{(H)}(f\bar f)$ involves $Tr_H$ rather
than the $Tr_G$ in $S_1(f\bar f)$ of equation (2.4),
a factor of $N_f$ appears. Note that had we introduced the
special parameterization of the gauge fields in the fermionic formulation
of $QCD_2$, we would have arrived at the same action after decoupling the
fermionic currents from the gauge fields, by performing chiral rotation
and then bosonizing the free fermions. Equation \mishwzwh\ was derived
without paying attention to possible renormalizations. The latter
will be treated in section 7.

At this point one may choose a gauge. A convenient gauge choice is
$\bar A=i\bar f\bpa\bar f^{-1}=0$. Notice that since the underlying
space-time is a plane this is a legitimate gauge. The gauge fixed action
can be written down using the BRST procedure, namely
$$\eqalign{S_{GF}=&S_{QCD_2}
+ S^{(gf)}+ S^{(gh)}=S_{QCD_2} + \delta_{BRST}(b \bar A)=\cr
=&S_{QCD_2} + Tr_H[B\bar A] + Tr_H[b \bar D c]\cr}\eqn\mishgf$$
where $S_{GF}$, $S^{(gf)}$ and  $S^{(gh)}$ are, respectively,
the gauge fixed action, the gauge fixing term and the ghost action. The
$(b,c)$ fields are yet another $(1,0)$ ghost system and $B$ is a
dimension one auxiliary field, all in the adjoint representation of
$SU(N_C)$. The integration over $B$ introduces a delta function of the
gauge choice to the measure of the functional integral. In addition we
integrate over the ghosts $b$ and $c$.

It is interesting to note that the \Q\ action can be related to a
``perturbed" topological  ${H\over H}$ coset model. To realize this face
of \Q\ we parameterize $u$ as $ghle^{i\sqrt{4 \pi\over N_CN_F}\phi}$ and
rewrite \mishgf\ accordingly.\refmark{\FS}  The
Polyakov-Wiegmann relation\refmark{\PW} implies
$$\eqalign{S[ u] &= S[ g   h
  l] +
{1\over 2} \int d^2x\partial_\mu\phi\partial^\mu\phi\cr
S[  g    h   l]&=
S[  g ] + S[   l] + S[  h ] +{1\over 2\pi}
\int d^2x
Tr(  g^\dagger \partial_+  g   l\partial_-  l^\dagger
+   h^\dagger \partial_+  h
  l \partial_-  l^\dagger).\cr}
\eqno{\eq})$$
Since  $l$ is a
dimension zero field  with an associated zero central charge we have   $S[l]=0$
 and  thus
$$\eqalign{S_{GF}&=S_{N_F}(h)+S^{(H)}_{-(N_F+2N_C)}(f)+{i\over{2\pi}}\int
d^2z Tr_H[\bar\rho\pa\bar\chi+\rho\bar\pa\chi]\cr
&+S_{N_C}(g)+{1\over2\pi}\int d^2z[\pa\phi\bpa\phi]\cr
&+{m^2\over{2\pi}}\int d^2z\Tr_G:[f^{-1}ghle^{i\sqrt{4 \pi\over N_C
N_F}\phi}+e^{-i\sqrt{4\pi\over N_CN_F}\phi}l^{-1}h^{-1} g^{-1}f]:\cr
&+{1\over e_c^2}\int d^2z\Tr_H[(\bpa(f^{-1}\pa f))^2]\cr
}\eqn\mishwzwghl$$
It is now easy to recognize the first line in the action as the
action of ${SU(N_C)\over SU(N_C)}$ topological theory.

It is interesting to note that a WZW term $S_{-2N_C}(f)$ appears in the
action even without the introduction of quarks. We therefore digress to
an analysis of the pure YM theory in the formulation introduced above.


\chapter{ Two dimensional Yang-Mills theory }

 Pure Yang-Mills theory attracted recently much attention
along the lines of an underlying string theory \refmark{\Gross}. Here we
restrict our discussion to the 2D Minkowski or Euclidean space-time,
 where the rich structure of the
model on compact Riemann surface does not show up.
In terms of the parameterization introduced in equation \mishwzwh\ the
 gauge invariant action of the  pure YM theory is
$$\eqalign{S_{YM_2}=S_{-(2N_C)}(f\bar f)
&+{i\over{2\pi}}\int d^2z\Tr_H[\bar\rho\pa\bar\chi+\rho\bar\pa\chi] \cr
&+{1\over e_c^2}\int d^2z\Tr_H[F^2]\cr}\eqn\mishwzwha$$
Here again we remind the reader that the coupling constant undergoes a
multiplicative renormalization. This will be discussed in section 7.
Let us first discuss the corresponding equations of motion for $f$ and
$\bar f$,
$$\eqalign{\delta f: \ \ \  \bpa A-D\bar A+{2\over  m_A^2}D\bar DF
=(1+{2\over  m_A^2}D\bar D)F&=0 \cr
\delta \bar f: \ \ \ \pa\bar A-\bar DA-{2\over  m_A^2}\bar DDF
=-(1+{2\over  m_A^2}\bar DD)F&=0}\eqn\misheqm$$
where $D=\pa-i[A,\cdot]$, $m_A=e_c\sqrt{N_C\over \pi}$ and $\dAlemb
=2\pa\bar\pa$. In fact these two equations are identical, as
$[D,\bar D]F=0$. The equation is that of a massive gauge field with self
interaction. Note that in this approach, unlike  the equations that
follow from varying the action with respect to the gauge fields, one gets
two derivatives on F.
In deriving the above, it is convenient to remember that
$$\delta S_{WZW}(f)={1\over2\pi}Tr\{
(f^{-1}\delta f) \bar\pa (f^{-1}\pa f)\}.$$

The YM action equation \mishwzwha\ is obviously invariant under the original
gauge transformations
$$f\rightarrow fv(z,\bar z)
\qquad \bar f\rightarrow v^{-1}(z,\bar z)\bar f\eqn\mishdels$$
with $v\in SU(N_C)$. In addition the action is invariant separately under
the holomorphic and anti-holomorphic ``color" transformations
$$f\rightarrow  u(\bar z) f\qquad \bar f\rightarrow  \bar f w(z)
\eqn\mishdels$$ where $u,w \in SU(N_C)$.
These are ``spurious" transformations since they leave $A$ and $\bar A$
invariant.
The corresponding holomorphic and anti-holomorphic color currents are
$$\eqalign{\bar J^s&=-{N_c\over\pi}[i(f\bar f)\bar\pa(f\bar f)^{-1}
-{2\over m_A^2} f\bar DFf^{-1}]\cr
J^s&=-{N_c\over\pi}[i(f\bar f)^{-1}\pa(f\bar f)
+{2\over m_A^2}\bar f^{-1}DF\bar f]\cr}\eqn \conlaw$$

The gauge fixed ($\bar f=1$) action takes the form
$$\eqalign{S_{YM_2}=S_{-(2N_C)}(f)&+{1\over e_c^2}\int d^2z\Tr_H[(\bpa
(f^{-1}\pa f))^2]\cr
&+{i\over{2\pi}}\int d^2z\Tr_H[\bar\rho\pa\bar\chi+\rho\bar\pa\chi
 ]\cr}\eqn\mishwzwgf$$
As is expected the equation of motion at present is just that of equation
\misheqm\ after setting $\bar A=0$. Naturally, the action now lacks gauge
invariance, nevertheless,  it is invariant under  the following residual
holomorphic transformations
$$f\rightarrow  u(\bar z) f\qquad  f\rightarrow   f w(z)
\eqn\mishdels$$
with the corresponding  holomorphic and anti-holomorphic currents
$$J_G=-{N_C\over\pi}[A+{2\over m_A^2} D(\bpa A)] \qquad
\bar J_G=-{N_C\over\pi}[\t A+{2\over m_A^2} \bar{\t D}(\pa\tilde A)]
\eqn\mishjc$$  and $\t A=if\bpa f^{-1}$. Notice that in spite the similar
structure, $\t A$ is not related to $\bar A$ which was set to zero. To
better understand the physical picture behind these currents we defer
temporarily to the abelian case.
\chapter{ Schwinger model revisited}

Since in the pure Maxwell theory there is no analog to
the $(-2N_C)$ level WZW term of equation \mishwzwha, we study instead the
Schwinger model in its bosonized form,
$$ S_{(Sch)}={1\over2\pi}\int d^2z[\pa X\bpa X-\sqrt{2}\pa X\bar A
+\sqrt{2}\bpa
XA+{\pi\over e^2} (\pa \bar A-\bpa A)^2]\eqn\mishsch$$
In analogy to the change of variables in the non-abelian case,
we now introduce      the following parameterization of the gauge fields
$A=\pa \Cf,\   \bar A=\bpa \bar\Cf. $ In terms of these  fields the
action takes the form
$$S_{(Sch)}={\int{d^2z\over2\pi}\{[\pa X\bpa X
-\sqrt{2}X\pa\bpa(\Cf-\bar\Cf)+
{\pi\over e^2}[\pa\bpa(\Cf-\bar\Cf)]^2+i[\bar\rho\pa\bar\chi+
\rho\bar\pa\chi]\}}\eqn\mishCf$$
In the gauge $\bar A=0$ and after the field redefinition
$\t X=X+{1\over \sqrt{2}}\Cf$, the action is
decomposed into decoupled sectors
$$\eqalign{S_{(Sch)}&=S(\t X)+S(\Cf)+S_{(ghost)}\cr
S(\t X) &= {1\over2\pi}\int d^2 z [\pa\t X\bpa\t X]\qquad
S(\Cf)= {1\over 4\pi}\int d^2 z\{{2\over\mu^2}
[\pa\bpa\Cf]^2
-\pa\Cf\bpa\Cf\},\cr} \eqn\mishtCf$$
where $\mu^2={e^2\over\pi}$.
The corresponding equations of motion are
$$\pa\bpa[1+{2\over\mu^2}\pa\bpa]\Cf=0,\qquad\pa\bpa\t X=0\eqn\mishEM$$
The invariance  under the chiral shifts
$\delta\Cf=\epsilon(\bar z)$ and  $\delta\Cf=\epsilon(z)$ are
generated by the holomorphically  conserved currents
$$J_G=\pa\Cf+{2\over\mu^2}\pa\bpa\pa\Cf,\qquad
\bar J_G=\bpa\Cf+{2\over\mu^2}\pa\bpa\bpa\Cf.\eqn\mishJJ$$
To handle this type of ``hybrid" current  we suggest the following
decomposition of the massless and massive modes $\Cf=\Cf_1+\Cf_2$ with
$$\pa\bpa\Cf_1=0\qquad[2\pa\bpa+\mu^2]\Cf_2=0\eqn\msihFF$$
In the holomorphic quantization
$$\Pi={\delta{\cal L}\over\delta(\pa\Cf)}=-{1\over \pi\mu^2}\bpa(\pa\bpa
+{\mu^2\over4})\Cf={1\over4\pi}\bpa(\Cf_2-\Cf_1).\eqn\mishPi$$
A unique solution to the commutation relations
$[\Cf(z,\bar z),\Pi(w,\bar w)]_{z=w}=i\delta(\bar
z- \bar w)$, $[\Cf,\Cf]=0$ and $[\Pi,\Pi]=0$ is
$$\eqalign{[\Cf_1(z,\bar z),\Cf_1(w,\bar w)]_{z=w}=&\pi i\epsilon(\bar
z-\bar w)\cr
[\Cf_2(z,\bar z),\Cf_2(w,\bar w)]_{z=w}
=&-\pi i\epsilon(\bar z-\bar w)\cr
[\Cf_1(z,\bar z),\Cf_2(w,\bar w)]_{z=w}
=&  0 \cr
[\t X(z,\bar z),\t X(w,\bar w)]_{z=w}
=&-\pi i\epsilon(\bar z-\bar w)\cr
}\eqn\mishcom$$
where $\epsilon$ is the standard antisymmetric step function.
Notice that the massless degree of freedom has commutation relations
which correspond to a negative metric on the phase space.
These relations can also be translated to the following OPE's
(choosing the part $\Cf_1(z)$ of $\Cf_1$)
$$\eqalign{\Cf_1(z)\Cf_1(w)&=log(z-w)\cr
\Cf_2(z,\bar z)\Cf_2(w, \bar w)&=-log|(z-w)|^2+ O(\mu^2|z-w|^2)\cr
\Cf_1(z)\Cf_2(w, \bar w)&=O(z-w)}\eqn\mishOPE$$
It is thus clear that the model is invariant under a $U(1)$ affine Lie
algebra of level $k=-1$ since $J_G(z)J_G(w)={1\over(z-w)^2}$,
as $J_G$ has no contribution from $\Cf_2$ and $J_G=\pa\Cf_1$ (no contribution
from $\Cf_2$).

The physical states of the model have to be in the cohomology of the BRST
charge. Due to the fact that the current is holomorphically (and the other
anti-holomorphically) conserved, it follows that the same property holds
for the BRST charge, and thus the space of physical states is an outer
product of the cohomology of $Q$ and $\bar Q$.
The latter are given by\refmark\us
$$\eqalign{Q=&\chi J =\chi(i\pa \t X+\pa \Cf_1), \cr
\bar Q=&\bar\chi\bar J={\bar\chi}(-i\bpa\t X+\bpa\Cf_1).}
\eqn\mishQQ$$

Expanding the fields $i\pa\t X$ and $\pa\Cf_1$ in terms of the
Laurant modes  $\t X_n$ and $(\t\Cf_1)_n$ with
$[X_n,X_m]=n\delta_{n+m}$ and $[(\Cf_1)_n,(\Cf_1)_m]=n\delta_{n+m}$
we have
$$Q=\sum_{n}\t\chi_n\[\t X_{-n}-i(\Cf_1)_{-n}\].$$
Since $J_0=\{Q,\rho_0\}$, physical states have to have zero eigen-value of
 $J_0$.
The general structure of the states in the $\Cf_1,\t X, \rho, \chi$
Fock space are as follows
$$(\t X_n)^{n_X}({\Cf_1}_m)^{n_f}(\chi_k)^{n_\chi}(\rho_l)^{n_\rho}
|vac\>$$ where obviously ${n_\chi}$ and ${n_\rho}$  are either 0 or $1$.
It is straightforward to  realize that only the vacuum  state and
states of the form $(\t X_0)^{n_X}({\Cf_1}_0)^{n_f}$ are in the BRST
cohomology. Recall that being on the plane we exclude zero modes
and thus only the vacuum state remains.
 Since there is no constraint on the modes of $\Cf_2$,
the physical states  are built solely
of $\Cf_2$ which are massive modes.
 This result is identical to the well
known solution of the Schwinger model.

\chapter{Back to the YM theory}
Equipped with the lesson from the Schwinger model we return now to the YM
case and introduce a decomposition of the group element $f$ so that again
the gauge currents obey an affine Lie algebra.
Let us write $f=f_2f_1$ which implies that
$$A=if^{-1}\pa f=if_1^{-1}\pa f_1+if_1^{-1}(f_2^{-1}\pa f_2)f_1\equiv
J_1+J_2\eqn\mishJJ$$
With no loss of generality we take $\bpa f_1=0$ implying also $\bpa J_1=0$.
 Inserting these expressions into $J_G$ of equation \mishjc\  one finds
$$J_G =-{N_C\over\pi}[J_1 +J_2 +{2\over m_A^2}(\pa\bpa J_2 +i[\bpa
J_2, J_1+J_2])]. \eqn\msihJC$$
If one can consistently require that
$$J_2+{2\over m_A^2}(\pa\bpa J_2+i[\bpa J_2,J_1+J_2])=0\eqn\Jtt$$
then, in a complete analogy with the abelian case,
$J_G=-{N_c\over\pi}J_1$.
The latter is an affine current of level $k=-2N_C$. One can in fact show
that \Jtt\ can be assumed without a loss of generality. $\bpa J_G=0$
implies that $J_2+{2\over m_A^2}(\pa\bpa J_2+i[\bpa
J_2,J_1+J_2])=u(z)$, where $u(z)$ is some
holomorphic function. We then introduce the  shifted
currents $\t J_2=J_2-u(z)$, $\t J_1=J_1+u(z)$. Now $\bar\pa\t J_1=0$ as does
 $J_1$,
 and $\t J_2$ obeys equation \Jtt\ with $\t J_1$ replacing $J_1$.
It is easy to check that the shifts in the currents correspond to
$f_1\rightarrow v(z)f_1$, $f_2\rightarrow
f_1v(z)^{-1}$  with $u(z)=i f_1^{-1} (v(z)^{-1}\pa v(z))f_1 $.

Note that the equation for $J_2$ involves  a coupling to $J_1$. This
is related to the
fact that, unlike the abelian case, one cannot write the action as a
sum of decoupled terms  which are functions of $J_1$ and $J_2$
separately.

Once the color current $J_G$ is expressed in terms of the
holomorphic current $J_1$, the analysis of the space of physical
states is directly related to that of the topological \G model
at $k=0$.\refmark\us\ The physical states have to be in the \co\ of
the BRST charge, which corresponds to the following holomorphically
conserved BRST current
$$ Q(z) = \chi^a (J_G^a +\half J_{gh}^a) =
-{N_C\over\pi}\chi^a(  [A+{2\over m_A^2}D(\bpa A)]^a +{i\over 2}
f^a_{bc} \rho^b\chi^c).\eqn \mishbrst$$
An anti-holomorphic BRST current $\bar Q(z)$ determines the
condition for physical states in the analogous manner to  $Q$. From here
on we restrict our description to the latter.
 We define now the zero level affine Lie algebra current
$$J^a_{(tot)}   =J_G^a  +J_{(gh)}^a= J_G^a +i\fcr \eqn\mishbJ$$
and the $c=0$ Virasoro
generator $ T$
$$ T(z) =  -
{1\over N_C }:J_G^a J_G^a: +\r  {} a \pa \c   {} a. \eqn\mishbT$$
as well as dimension (2,0) fermionic current
$$G=-{1\over 2N_C}\rho_a J_G^a,$$ and realize the existence of the
 ``topological coset algebra"\refmark{\IsRa,\Ber}
$$\eqalign{  T(z) =\{ Q, G(z)\},\ \ \   Q(z) =&\{ Q ,
j^\#(z)\},\ \ J^a_{(tot)}=\{ Q, \r {} a(z)\},\cr
\{Q,Q(z)\}=0,\ \ \ \ \ \{G,G(z)\}&\equiv W(z),\cr
W(z)=\{Q,U(z)\},\ \ \ \ \ \ \ \ \ &[W,W(z)]=0,\cr}
\eqn\mishbalgebra$$
where  $J^\#= \c a {}
\r {} a $ is ``ghost number current",
$$W(z)={1\over 4N_c} f_{abc}J_G^a\rho^b\rho^c + \pa\rho^a\rho_a$$
 and $$U={1\over 12N_c}f_{abc}J_G^a\rho^b\rho^c.$$
A direct consequence is that any physical state has to obey
$${J_{(tot)}}_0^0 |phys>=0,\qquad
L_0|phys>=0,\qquad W_0|phys>=0,\eqn\mishLJ$$
where $\jt n i$, $\t L_n$ and $W_n $ are the Laurent modes of $\jt {} i$
the Cartan sub-algebra currents, $ T$ and $W$ respectively.  In fact
the BRST cohomology of the present model is a special
case of the set of models of refs. [\us,\uss]. We therefore refer the
reader to those works and present here only  the result.
On the plane where no ghost zero modes are allowed, the only state in the
cohomology is the  zero ghost number  vacuum state of $J_1$.

 This state can be tensor-product with  oscillators of the massive modes  of
$J_2$.
Unlike the abeilan case, $J_G$ does not commute with $J_2$ so
that in general the $J_2$ modes are not obviously
 in the BRST \co.
However, there is no reason to believe that  all the $J_2$ modes will be
excluded by the BRST condition.
 Those $J_2$ modes that remain
are by definition color singlets.
This result contradicts previous results on $YM_2$.  Usually  one
believes that pure gluodynamics on the plane is an empty theory
since all local degrees of freedom can be gauged away.

\chapter{ An Alternative Formulation}
To get a better understanding of the subtleties
of the Yang Mills
theory when expressed in terms of  $A=if^{-1}\pa f$, $\bar A=i\bar f\bpa\bar
f^{-1}$, and for future application,
we compare now with another formulation of the theory.

Consider the following functional integral

$$ \eqalign{Z=&\int DA D\bar A DB e^{iS(A,\bar A, B)}\cr
S=& -\int d^2 z Tr_H[ {1\over e_c} FB + {1\over 4} B^2]\cr}\eqn\mishSBF $$
where $B$ is a pseudoscalar field in the adjoint representation.
Obviously the integration over $B$ produces the
usual $Tr[F^2]$ action. It is also easy to realize
that the action is invariant
 under the ordinary gauge symmetry  provided that $\delta B=i[\epsilon,B]$.
In terms of the $f$ variables after imposing the gauge $\bar f=1$ one finds
 $$S_{YM_2}=S_{-(2N_C)}(f)+\int
d^2z\Tr_H[( {i\over e_c}(f^{-1}\pa f)\bpa B) -{1\over 4}B^2] +S^{(gh)}
\eqn\mishBf$$
where $S^{(gh)}={i\over{2\pi}}\int d^2z\Tr_H[\bar\rho\pa\bar\chi
+\rho\bar\pa\chi ].$
One should again bear in mind that the coupling constant undergoes a
multiplicative renormalization. This will be discussed in the next section.
Using the results of ref. [\PW] we get
$$\eqalign{S_{YM_2}(B)=&\Gamma_{2N_C}(B)-{1\over 4}Tr_H[B^2] \cr
&-S_{(2N_C)}(vf)+S^{(gh)}\cr}\eqn\mishSBb$$
where $\Gamma_k(B)=S_k(v)$ with ${1\over e_c}\bpa B={2N_C \over2\pi}
(iv\bpa v^{-1})$.
The second line in \mishSBb\ is a $c=0$ ``topological system".
Since the underlying Minkowski space-time does not admit zero modes we can
safely
integrate over the corresponding fields.
We can further
 pass   from functional integrating  over $B$ to $iv\bpa v^{-1}$. This
involves the insertion of $det{\bar D\over \bar\pa}$ which will introduce a
$\Gamma_{-2N_C}(B)$ term with no additional ghost terms.
The functional integral \mishSBF\ thus takes the final form
$$ Z= \int D [v]  e^{-i\({1\over 4}\int d^2zTr_H[B^2]\)}. \eqn\mishBFF$$
It is thus clear that the in the present formulation there is no trace of the
massive
``physical modes" discussed in the previous section.

\chapter { The resolution of the puzzle}
Encouraged by  the result of the last section, we proceed now to reexamine the
steps
that led to the unexpected massive modes in  the pure $YM_2$ theory.
In particular, we would like to check whether
in addition to the implementation of
proper determinants
there is no  coupling  constant renormalization
that has to be invoked  when passing to
the quantum theory expressed in the $f$ variables.
For this purpose we turn on again the matter degrees of freedom.
We introduce $N_F$
quarks in the fundamental color representation and explore the
behavior of the
system in the limit  $N_F\rightarrow 0$.
Recall that the action of this model is given in equation \mishwzwghl.
Starting actually from equation \mishwzw, taking the massless limit, writing
$A$ in terms of the $f$ in the action but still with $A$ as
an integration variable,
 and using the formulation presented
in the previous section, the path integral of
 the colored degrees of freedom now reads
$$\eqalign{Z^{(col)}&=\int[DA][DB][Dh] e^{iS^{(col)}} \cr
S^{(col)}&=S_{N_F}(h)+ {N_F\over 2\pi}\int d^2z\Tr_H[ h\bpa
h^{-1}f^{-1}\pa f]\cr
&+\int d^2z\Tr_H\[\({i\over e_c}(f^{-1}\pa f)\bpa B\)-
{1\over 4}B^2\] \cr}\eqn\mishRP$$
where we have also gone from $u$ to $h$ as in section 2.

It was found out \refmark\Kut that  quantum consistency imposes
finite renormalization on
the coupling constant of  the current-gauge field interaction.
 This renormalization is expressed in the
following equality,
$$ \eqalign{Z(\bar J) \equiv &\int DA e^{i\[S_k(f)+ {1\over
2\pi} \int d^2z\Tr_H[ i(f^{-1}\pa f) \bar
J\]}\cr  =& \int D f e^{i[S_{k-2N_C}(f)+ {e(-k)\over 2\pi}\int
d^2z\Tr_H[  (f^{-1}\pa f )\bar J] }\int D(gh)e^{iS^{(gh)}}\cr
=& e^{i\Gamma_{-k+2N_C}\[\({e(-k)\over -k+2N_C}\)\bar
J\]}\int D(gh)e^{iS^{(gh)}}.\cr}\eqn\mishKut$$
 where $k$ is an arbitrary level and $\Gamma_k(L)=S_k(w)$ for
$L =iw\bpa w^{-1}$.
The renormalization factor $e(k)$ has to satisfy
 ${e(-k-2N_C)\over e(k)} = {k\over k+2N_C}$. In addition
it is clear from  equation \mishKut\  that it has to be singular at the
origin.
In ref. [\Kut],  $e(k)$ was taken to be  $e(k)=\sqrt{k+2N_C\over k}$.
Implementing this renormalization in  our case equation \mishRP\ takes the form
$$\eqalign{S^{(col)}=&S_{N_F}(\tilde
h)+ S_{-(N_F+ 2N_C)}(f) \cr
+&\int d^2z\Tr_H\[\( i \sqrt{{N_F+ 2N_C\over N_F}}
{1\over e_c}(f^{-1}\pa f)\bpa B\) -{1\over 4}B^2\]\cr +&
 S^{(gh)}\cr}\eqn\mishRPa$$
where $\tilde h = fh$.
After integrating the auxiliary field $B$  the action becomes
$$\eqalign{S^{(col)}&=S_{N_F}(\tilde h)+ S_{-(N_F+ 2N_C)}(f)
 +\int d^2z\Tr_H\[\( {N_F+ 2N_C\over  e_c^2 N_F}\)\]
[\bpa(f^{-1}\pa f)]^2]\cr &+ S^{(gh)}} \eqn\mishRPp$$
It is now straightforward to realize that the equation of
 motion which follows from
the variation with
respect to $f$ is that of equation \misheqm\ where now $m_A=
e_c\sqrt{N_F\over 2\pi}$.  Thus, the coupling constant
renormalization  turned the massive modes into massless  ones
in  the case of pure YM theory ($N_F=0$). Notice that to
reach  this  conclusion it is enough to use the fact that  $e(k)$
has to be singular at $k=0$ and the explicit expression of $e(k)$
is  really not needed. Following the arguments presented in section
5, it is clear that these states that became massless  are not in
the  BRST cohomology and thus not in the physical spectrum.

A somewhat similar derivation of  the triviality of the model in
the $N_F=0$ limit  is the following. We integrate in equation
\mishRPa\ over the ghost fields and over $f $
 using again the coupling constant
renormalization and find
$$ Z= \int D[v]  e^{-\{iS_{N_F}(v) +{e_c^2 N_F^2 \over 4}\int
d^2zTr_H[B^2(v)]\}}. \eqn\mishBFFa$$
It is now clear that
 the action
vanishes at $N_F=0$ and hence again, on trivial topology, the
theory is empty.  Notice,
however, that the implementation of renormalization modifies also the
result of the previous  section.

The final conclusion is that in both methods
 one finds  that indeed the pure YM theory
 has an empty space of  physical states as of course is implied by  the
original formulation in terms of $A$. We have demonstrated that in this
formulation it follows only after taking subtleties of renormalization
into account.

   \chapter{ On bosonized  \Q}
To resolve the puzzle of the YM theory  we were led
to analyze the
color and flavor
sectors of \Q. The full bosonized \Q\  includes in addition
the baryon number degrees of freedom.
The  corresponding action
 is  given by equations \mishwzwh, \mishgf\ or by equation
\mishwzwghl.
In the past the low lying baryonic spectrum  in the
strong coupling limit ${m_q\over e_c}\rightarrow 0$ was extracted using a
semiclassical quantization.\refmark{\DFS,\FS} In addition the ``quark"
structure of the model was analyzed in ref.[\FHK].
In the present work our analysis was based on
switching off the mass term, $m_q=0$.
This limit cannot be treated by the semi-classical approach,
as the soliton solution is not there for $m_q=0$.
In our case here one finds a decoupled $WZW$
action for the flavor degrees of freedom $S_{N_C}(g)$ and a
decoupled free field action for the baryon degree of freedom,
in addition to the action of the colored degrees of freedom which
is given in equation \mishRPp\ or equation \mishBFFa.
 The general structure of a physical state in this case is that
of a tensor product of $g$ and $\phi$ with the colored degrees of
freedom $f$, $h$ and the ghosts.
 The structure of \Q\ which
emerged from the semiclassical quantization \refmark{\DFS,\FS}
for $m_q\not=0$ involves $g$ and $\phi$ only. In
our case here
the $f$ colored
degrees of freedom acquire  mass $m_A=
e_c\sqrt{N_F\over 2\pi}$
 while the $h$ degrees of freedom remain
massless. In the  limit
$e_c\rightarrow\infty$ the $f$ degrees of freedom decouple.
 It is
thus clear that one has to introduce the mass term which couples the
three sectors. The massless limit of \Q\ can then be derived by
taking the limit $m_q \rightarrow 0$
after solving for the physical states.
Indeed, it was shown \refmark\DFS  in the limit of $e_c\rightarrow \infty$
that turning on $m_q\neq 0$ results in a hadronic spectrum where the
flavor representation and the baryon number were correlated.
The analysis of the spectrum of the massive multi-flavor \Q\
in the approach of this work remains to be worked out.

\chapter{ Summary and Discussion}
In this work we have analyzed 2D YM and QCD theories using a special
parametrization of
the gauge fields in terms of group elements. In the $m_q=0 $
case it enabled us
to decouple the matter and gauge degrees of freedom. However, this
formulation led, in a naive treatment, to unexpected massive modes.
Even though
we did not present a full solution of the theory we had reasonable
arguments
to believe that the BRST projection would not exclude these modes.
The fact that a similar approach to $QED_2$  reproduced the known results
of the
Schwinger model, enhanced the puzzling phenomenon.
Eventually, we showed that a coupling constant renormalization, which
was originally introduced in ref. [\Kut], renders the unexpected
massive modes
into massless unphysical states.  The benefit of this detective
work is the appearing
of ``physical" massive states in massless \Q. These
states of mass
$m_A=e_c\sqrt{N_F\over 2\pi}$  were actually discovered in ref.
[\Pat] using another approach.
The general task of a full
solution of \Q\ still remains to be worked out.

\ack{We are indebted to David Kutasov for pointing out to us the
renormalization factor that accompanies the change of variables
from A to $f$, as in equation \mishKut.
We would also like to thank M. Porrati and S. Yankielowicz
for useful conversations.}
\refout
\end
\bye